%% file: prosocialllms.tex
\documentclass[sigconf]{acmart}
\usepackage{enumitem}
\usepackage{array}

\AtBeginDocument{%
  }

\copyrightyear{2025}
\acmYear{2025}
\setcopyright{acmlicensed}\acmConference[IUI '25]{30th International Conference on Intelligent User Interfaces}{March 24--27, 2025}{Cagliari, Italy} \acmBooktitle{30th International Conference on Intelligent User Interfaces (IUI '25), March 24--27, 2025, Cagliari, Italy} \acmDOI{10.1145/3708359.3712149} \acmISBN{979-8-4007-1306-4/25/03}

\begin{document}

\title{Simulating Cooperative Prosocial Behavior with Multi-Agent LLMs: Evidence and Mechanisms for AI Agents to Inform Policy Decisions}

\renewcommand{\shorttitle}{Multi-Agent Simulations of Cooperative Prosocial Behavior}

\author{Karthik Sreedhar}
\email{ks4190@columbia.edu }
\affiliation{%
  \institution{Columbia University}
  \city{New York}
  \state{New York}
  \country{USA}
}

\author{Alice Cai}
\email{acai@college.harvard.edu}
\affiliation{%
  \institution{Harvard University}
  \city{Cambridge}
  \state{Massachusetts}
  \country{USA}
}

\author{Jenny Ma}
\email{jm5676@columbia.edu}
\affiliation{%
  \institution{Columbia University}
  \city{New York}
  \state{New York}
  \country{USA}
}

\author{Jeffrey V. Nickerson}
\email{jnickers@stevens.edu}
\affiliation{%
  \institution{Stevens Institute of Technology}
  \city{Hoboken}
  \state{New Jersey}
  \country{USA}
}

\author{Lydia B. Chilton}
\email{chilton@cs.columbia.edu}
\affiliation{%
  \institution{Columbia University}
  \city{New York}
  \state{New York}
  \country{USA}
}

\begin{abstract}
Human prosocial cooperation is essential for our collective health, education, and welfare. 
However, designing social systems to maintain or incentivize prosocial behavior is challenging because people can act selfishly to maximize personal gain.  
This complex and unpredictable aspect of human behavior makes it difficult for policymakers to foresee the implications of their designs.
Recently, multi-agent LLM systems have shown remarkable capabilities in simulating human-like behavior, and replicating some human lab experiments. 
This paper studies how well multi-agent systems can simulate prosocial human behavior, such as that seen in the public goods game (PGG), and whether multi-agent systems can exhibit ``unbounded actions'' seen outside the lab in real world scenarios. We find that multi-agent LLM systems successfully replicate human behavior from lab experiments of the public goods game with three experimental treatments - priming, transparency, and varying endowments. Beyond replicating existing experiments, we find that multi-agent LLM systems can replicate the expected human behavior when combining experimental treatments, even if no previous study combined those specific treatments. Lastly, we find that multi-agent systems can exhibit a rich set of unbounded actions that people do in the real world outside of the lab -- such as collaborating and even cheating. In sum, these studies are steps towards a future where LLMs can be used to inform policy decisions that encourage people to act in a prosocial manner.
\end{abstract}



\maketitle

\input{1_intro}
\input{2_related_work}
\input{3_4_reordered}
\input{5_discussion}

\input{6_conclusion}

\bibliographystyle{ACM-Reference-Format}
\bibliography{prosocialllms}

\end{document}

%% file: 1_intro.tex
\section{Introduction}

Human prosocial cooperation is essential for our collective health, education, and welfare, but cooperation is also difficult to maintain. 
When everyone pays taxes, societies can collectively benefit from public goods like roads, safety nets, and schools. However, individuals who do not contribute get to keep their wealth and still benefit from public goods. 
This "freeloading" behavior can eventually erode public trust and lead to a tragedy of the commons. 
Psychologists study this behavior in the lab using games like the public goods game (PPG) where a group of 3 or more people all decide how much to donate to a public fund. Public funds are multiplied then redistributed equally. If everyone donates, everyone is better off. However, if someone decides not to donate, they benefit more than others - they receive the public funds and get to keep their initial wealth. Eventually, this freeloading behavior discourages everyone else from donating, and public funds are emptied. 
Encouraging cooperation is thus important and psychologists have discovered many factors that affect prosocial behavior such as priming (positive framing), transparency (social pressure), and varying endowments (inequality).  

However, human behavior in the lab is different than in the wild. In the lab, people’s choices and actions are bounded: they only get to choose from a small set of options, such as a choice from a menu of how much to donate. But in the wild, choices are unbounded. Humans exhibit more complex behaviors including lying, cheating, persuasion, and gaming the system. These are often the behaviors that cause programs and policies to fail. They’re also very difficult to foresee. Studying human behavior in the wild using experiments is very challenging. It requires large numbers of people interacting, creating a control group or a matched sample to compare to, and numerous runs a to achieve statistical significance. This is almost never done, and thus policies affecting human cooperative behavior are rarely informed by experimental evidence.

Multi-agent LLM systems have the potential to simulate complex human behavior and interactions. 
They exhibit human-like behavior \cite{park2023generativeagentsinteractivesimulacra}, including doing actions they were not explicitly told to do (emergent behavior), thus displaying the potential for unbounded actions. 
Beyond human-like behavior, LLMs and multi-agent LLM systems have also shown a capacity to replicate actual human behavior seen in lab studies. This includes 70 psychology experiments \cite{aher2023usinglargelanguagemodels, hewitt2024predicting}, and multi-player economic games with strategic thinking about what other players will do, such as the 
multi-turn Ultimatum Game \cite{sreedhar2024simulatinghumanstrategicbehavior} and auctions \cite{manning2024automated}. 
For a simulation to be useful, it does not necessarily have to replicate human behavior exactly, but it does have to exhibit a rich enough set of actions consistent with behaviors that are observed, and provide evidence that the results can generalize to previously unseen situations.
To lay the groundwork for a future where policy makers can ask LLM-agents to simulate human behavior in response to policy changes, we address the following  questions:
\begin{itemize}
    \item[(1)] Do multi-agent LLM systems understand human cooperative behavior well enough to simulate it, even in a lab setting?
    \item[(2)] Are multi-agent LLM systems simply echoing the results from previous papers, or can effects transfer across papers and apply more broadly?
    \item[(3)] To what extent are multi-agent LLMs complex enough to simulate the rich set of unbounded options, actions, and interactions people do in the real world, outside of the lab? 
    \item[(4)] To see the complex interactions observed in the real world, what simulation mechanisms do we have to add to the set up of the system?
\end{itemize}

Our experiments show that multi-agent LLM systems replicate findings of three lab studies of human cooperative prosocial behavior, showing that LLM-agents are consistent with humans in their responses to priming, transparency and varying endowments. The simulation accurately replicated the direction of the effect (positive or negative), but often exaggerated the effect size. In addition to replicating the findings of previous studies, we show that multi-agent LLM systems can apply insights from other games to the PGG. Specifically, we demonstrate that two priming effects previously studied in other games — but not tested in the PGG — also have a similar impact on behavior in the PGG. This indicates that multi-agent LLM systems are not simply echoing existing research but have a broader ability to simulate psychological mechanisms across different games, situations, and simulations.

We also show that for two real world situations, we do see complex, unbounded behavior consistent with anecdotal evidence. For a classroom scenario with a teacher testing various late policies, we find that when late policies are strict and assignments are hard, we can enable the behavior of cheating to emerge. For a shopping center parking lot scenario, we find that LLM-agents can be affected by external conditions like humans in returning or not returning their shopping cart. For both scenarios, simple mechanisms had to be added to the simulation to see these effects. For cheating, the simulation had to include communication channels like a private student room to initiate the communication needed for a behavior like cheating. For agents to not return their cart, their provided information had to allude to what was at stake as a result of their external condition.

Altogether, we conclude that multi-agent LLM systems show remarkable promise towards simulating human prosocial behavior.  
For broader real-world scenarios, we find that there remain additional mechanisms to identify in order for multi-agent LLM systems to simulate cooperative behavior, including (1) creating communication mechanisms that allow agents to coordinate behavior privately and (2) make sure the stakes (or incentives) driving behavior are clear. 
These are far from the only mechanisms needed, and there is much future work to discover more mechanisms, but there is reason to believe that these mechanisms
will be broadly applicable across domains. 
Our simulations show that in the future, there is a potential for multi-agent LLM simulations to be useful in informing policy makers. Currently, LLM-agents can simulate the general direction of human behavioral effects, although they struggle to accurately capture the magnitude. While fine-tuning may improve this, it likely never perfectly replicates human behavior. 
Nevertheless, the ability to simulate directional trends can still help policymakers anticipate potential policy outcomes. Given the complexity and high social stakes of such decisions, even preliminary simulations could provide a valuable framework for exploring possible scenarios and guiding decision-making.

%% file: 2_related_work.tex
\section{Related Work}

\subsection{The Public Goods Game}

The public goods game (PGG) is a particularly important case of human cooperative prosocial behavior studied by psychologists and economists. In the PGG, a set of three or more players are each given an endowment of money, of which they choose a portion of to voluntarily contribute in-private to a public pool. The public pool will then be multiplied by a factor, say 2, then redistributed evenly amongst the players. In a game with four players, if every player has a \$20 endowment and contributes half of their endowment, \$10, each player will receive a payoff of \$30: \$10 from the amount of their endowment they kept and \$20 from the payoff from the public pool. If each player donates nothing, they all simply keep their initial endowment of \$20 without any additional payoff. In the PGG, working cooperatively can be mutually beneficial. However, if a player decides not to contribute to the public pool, the player can increase the payoff — in the example previously discussed, if one player decides not to contribute to the public pool, the payoff is \$35 — \$20 from the endowment the player kept, and \$15 from the payoff from the public pool. Hence, not contributing to the public pool can result in a greater payoff for an individual, despite contribution to the public pool by \textit{someone} being necessary for there to be any additional payoff for players. This game mimics situations involving prosocial behavior in the real-world, such as taxation — when paying taxes, people are forced to contribute some of their own money to a public pool to develop infrastructure and services for the broader community. People can even try to avoid paying taxes through tax havens or loopholes and freeload off public goods.

Lab experiments with human subjects show that in single-round (one-shot) PGGs, people do typically contribute a portion of their endowment — on average, 50\%. When the game is played for multiple rounds, the amount that players contribute decreases each round, until it is effectively zero \cite{labexperiments}. However, introducing variations to the game can increase or decrease prosocial behavior. Priming participants by presenting the game under different names \cite{Eriksson_Strimling_2014} or having participants primed with words alluding to cooperation before playing the PGG \cite{drouvelis2015priming} increases their average contribution amount. Introducing transparency of contributions also increases average contributions — having participants write their contribution on a blackboard after each round \cite{rege2004impact} or announce their contributions publicly \cite{transparencytwo} increases the average contributions made in the PGG. Finally, varying the endowments that players have to begin with in the PGG also has an effect on average contributions — in experiments with individuals given low (\$20), medium (\$50), and high (\$80) endowments, individuals with the low and medium endowments contributed the same amount in games where endowments were the same as in games where endowments were varied. However, individuals with the high endowment contributed much less in games where endowments were varied \cite{HARGREAVESHEAP20164}.

These effects are not limited to the PGG and can be seen in select real-world scenarios. In fundraisers, audiences may be primed to cooperate towards a common goal. In many settings, pledges are made publicly (that is, transparently), which increases donations \cite{Bhati2020-xw, Oppenheimer2011-zh}. Similar strategies are used by policy makers seeking to levy taxes for a bridge or other local public goods \cite{parks2013cooperation}. Furthermore, lower class individuals have been demonstrated to be more prosocial than their higher class counterparts in studies of generosity, charity, trust, and helpfulness \cite{piff2010having}. Overall, although free-loaders may be able to benefit off of public goods — whether in the PGG or in the real world — there are strategies that policy makers can use to increase prosocial behavior.

\subsection{Replicating Bounded Scenarios with Human Subjects using LLM Simulations}

Prior research in psychology, economics, and artificial intelligence has compared results from LLM simulations to experimental data from studies with human subjects, finding that outcomes in LLM simulations generally mirror those from the real world. Using a single LLM, past work has been able to replicate results from economic and psychology experiments with humans, including the Ultimatum Game, Garden Path Sentences, Milgram Shock Experiment, and Wisdom of Crowds \cite{aher2023usinglargelanguagemodels}. Single LLMs have also been able to replicate results from an archive of seventy social science experiments — both published and unpublished. In published studies, LLM simulation results had correlation coefficient of r = 0.85 with results from lab experiments with humans. In unpublished studies which could not have appeared in the LLM's training data, the correlation was also strikingly high (r = 0.90) \cite{hewitt2024predicting}. However, LLMs have been found to potentially miss nuances in human behavior in psychology experiments, potentially overestimating and giving false positives \cite{cui2024aireplacehumansubjects}. But overall, past work suggests that single LLMs do have a general sense of human behavior in tightly bounded experimental situations. 

Prior research has also used multi-agent LLM systems to replicate human behavior in economic and psychological settings. Prior work has used multi-agent LLM systems to simulate economic games like the Dictator Game, as well as agents' immediate responses to price increases or budget allocations, finding that it is possible to change LLM-agent behavior by giving them different endowments, information, and preferences \cite{horton2023largelanguagemodelssimulated}. Frameworks have also been presented to simulate scenarios such as negotiations, bail hearings, job interviews, and auctions, finding that LLM-agents are able to predict the signs of effects, but not the magnitude \cite{manning2024automated}. However, these studies often lack robust comparisons to human behavior in similar real-world situations. Multi-agent systems have also shown the ability to replicate human strategic behavior in the ultimatum game more accurately than single LLMs, as assessed by comparing simulation results to human experimental data  \cite{sreedhar2024simulatinghumanstrategicbehavior}. However, the ultimatum game is a single economic experiment that has been extensively studied and thus likely appears often in the training data of LLMs, leaving room for further exploration.



Past research has demonstrated that LLMs, and multi-agent LLM systems in particular, show promise in replicating human behavior in lab experiments where behavior is bounded. However, policymakers require simulations of unbounded scenarios - where people have nearly unlimited choices on ways to think, act, and behave. Our work begins by confirming that multi-agent LLM systems could replicate prosocial behavior in lab experiments, but then extends it into select real-world situations where agents had a more unbounded action space.\color{black}

\subsection{Multi-Agent LLM Systems}

Previously introduced multi-agent systems have varying capacities for allowing unbounded human behavior. Prior research has presented systems for simulating scenarios such as negotiations, bail hearings, job interviews, and auctions. In such systems, the action space is somewhat limited, with the system having rigid turn-taking guidelines and detailed information on how the simulation should proceed \cite{manning2024automated}. Similarly, ChatArena is a system which allows users to simulate turn-taking games: setups for tic-tac-toe, rock-paper-scissors, and chameleon are provided by default, while users can modify input information to simulate games such as the PGG. However, the system is again rigid in the actions agents can take and when agents can take such actions \cite{ChatArena}.

There are also multi-agent LLM systems in which LLM-agents are allowed more complexity and freedom in the ways they can "think" and act: these are referred to as "generative agents" \cite{park2022socialsimulacracreatingpopulated}. MetaAgents is a framework in which agents have four modules: perception, memory, reasoning, and execution. Using a simulated environment of a job fair, the research studies agents' abilities to create teams and workflows for collaborative tasks, finding that agents show promising results in simulation \cite{li2023metaagentssimulatinginteractionshuman}. Architectures have also been presented to simulate the inhabitants of a town. Agents have modules for observation, memory, reflection, planning, and reacting. Multiple locations can also be defined for agents to move between. In simulation, agents are found to demonstrate emergent behaviors, but not compared to observations of human actions explicitly \cite{park2023generativeagentsinteractivesimulacra}.

Past work has also explored coupling large language models with simulation models to better represent human behavior and decision-making \cite{Ghaffarzadegan_2024}, as well as using LLMs to mediate between end-users and simulations \cite{giabbanelli2024broadening}. However, there remains further room for exploration in identifying mechanisms within multi-agent systems themselves to achieve similar goals.



Whereas previous research shows that multi-agent LLM systems can simulate realistic human behavior, our work goes further showing that multi-agent LLM systems can simulate real observed human behavior in the lab. Additionally, we show that LLMs can simulate observed human behavior outside the lab, where there is an unbounded action space. For example, people can cheat.\color{black}

%% file: 3_4_reordered.tex
\section{Experimental Setup}

With the goal of better understanding the ability of multi-agent LLM systems to model prosocial human behavior, we simulate three different types of experiments: (1) simulations which aim to directly capture effects observed in a previous lab experiment with human subjects; (2) simulations which combine effects from multiple previous lab experiments with human subjects; and (3) select "in-the-wild" real-world scenarios we use as case studies towards identifying mechanisms necessary for simulations to display human-like behavior. For all experiments, we use OpenAI's GPT4 (model gpt-4o-mini), which has been shown to have the ability to interpret inherently human concepts such as equity and also scores  well on a variety of standardized tests, ranging from the Bar Exam to the GRE \cite{openai2023gpt4}. 

We specifically address the following three research questions:

\bigskip

\begin{itemize}[itemindent=0em]
  \item[\underline{\textbf{RQ1}}] Can multi-agent LLM system simulations replicate behaviors observed in PGG lab experiments with human subjects?
\begin{itemize} [leftmargin=0.15in]
\item[\underline{a.}] Does priming LLM agents via game name replicate the effect of priming humans via game name?
\item[\underline{b.}] Does introducing transparency in contribution between LLM agents replicate the effect of introducing transparency in contribution between humans?
\item[\underline{c.}] Does varying the endowments of LLM agents replicate the effect of varying endowments of humans?
\end{itemize}
\end{itemize}

%

  


\bigskip

\begin{itemize}[itemindent=0em]

  \item[\underline{\textbf{RQ2}}] Can multi-agent LLM system simulations transfer effects observed in non-PGG lab experiments to simulations of the PGG?
\begin{itemize}[leftmargin=0.15in]
    \item[\underline{a.}] Does priming LLM agents with a methodology used in non-PGG cooperation game lab experiments have the expected result in LLM simulations of the PGG?
\item[\underline{b.}] Does the effect of priming over time, observed to fade in non-PGG competition game lab experiments, hold in LLM simulations of the PGG?
\end{itemize}
\end{itemize}

  \bigskip
%



\begin{itemize}[itemindent=0em]
  \item[\underline{\textbf{RQ3}}] To what extent are multi-agent LLMs complex enough to simulate the rich set
of unbounded options, actions, and interactions people do in the real world, outside of the lab?
\begin{itemize}[leftmargin=0.15in]
    \item[\underline{a.}] To see the complex
interactions observed in the real world, what simulation mechanisms do we have to add to the set up of the system?
\end{itemize}
\end{itemize}


%



\bigskip

We expect generally positive results for the first two research questions, given that other work has shown the ability of LLMs to replicate lab experiments that are both published and unpublished \cite{hewitt2024predicting}, but still verify this in the context of the PGG towards understanding the abilities of multi-agent LLM systems to simulate prosocial behavior. The third research question is more open-ended - we use case studies of real-world scenarios towards identifying mechanisms required for simulations to show expected outcomes, a process which can be continued to be used by policy makers. As policy makers consider the specific scenarios which they need to simulate, they may realize and implement new mechanisms required to enable their simulation. However, we also expect that mechanisms will be generalizable to an extent across simulations of various situations.

\subsection{Multi-Agent Architecture}
For all studies, we adapted a previously introduced and open-source multi-agent system for simulating social emergent behavior \cite{GPTeam}. The system is implemented in Python and uses gpt-4o-mini. The input to the system is a JSON file describing the \textit{locations} and \textit{agents} in the simulation. The output is a series of logs for \textit{events} (things done by agents and witnessed by other agents). The input files and system we use for the simulations in our studies are publicly available \cite{prosocialGPTeam}.

The system architecture consists of a class structure defined as follows. At the highest level, the simulation is represented by a \textit{world} class consisting of 3 sub-classes: \textit{locations}, \textit{events}, and \textit{agents}. 

 Locations represent places that agents can go within the world.  Locations are specified by a name, description, and agents allowed within them. Agents can only talk to other agents or witness events in the same location as them. Events are created when agents take action, whether that be moving or speaking.  Events are specified by a timestamp, the agent that acted, the location it occurred, a description, and witnesses.

Agents are separate LLM instances that are used to represent "people" in the environment.  Agents are specified by a name, a private biography, a public biography and an initial plan. The private biography is information about the agent only known to the agent. The public biography is information about the agent that other agents also are aware of. The initial plan instructs the agent on what to do when the simulation starts. It is specified by a location (i.e., where it occurs), a description of what the agent should do, and a stop condition that specifies when the agent should stop executing on the plan. The initial plan is private - it is only known to the agent. 

Agents have subclasses \textit{plans} – specified by a description, location, and stop condition – and \textit{memories} – specified by a description, creation time, and importance score. Plans are created by agents as they witness events in simulation. All plans are specified in the same format as the initial plan. Memories are created each time an agent witnesses an event. When a memory is created, an importance score is also created simply by asking an LLM to generate one. When agents act, they first identify memories that are relevant and use them to maintain or adjust their current plan before executing on their action. 


Simulations can be run once the user specifies the locations and agents in the input file. The system executes simulation via \textit{agent loops}, which involve (1) agents \textit{observing} the events in their current location, (2) agents adding events to memory, (3) agents creating \textit{plans}, (4) agents \textit{reacting} as to whether or not to continue the current plan, (5) and agents \textit{acting}, carrying out plans. Agents \textit{speak} to communicate with one another.  The agent loop is what enables emergent behavior - agents are able to "reason" about their surroundings and events and then decide on new courses of actions. This type of architecture has been shown to be more accurate in replicating human behavior in social science/economics games than just prompting a single LLM \cite{sreedhar2024simulatinghumanstrategicbehavior}.

When implementing the PGG in this environment, we create 4-5 agents: 3-4 players, and one moderator. The private biography is used to provide agents with their unique priming or endowment that other agents should not know. There are typically two locations;  a game room where players are by default, and a contribution room where players make their contribution. Players are aware of the existence of these locations and can move between them. In every agent loop, the agent waits for the player before them to come back from the contribution room, then goes to the contribution room, tells the moderator their contribution, the moderator records it to his memory, and the player agent leaves. When all the players are done, the moderator comes out and announces the game outcome. 

\color{black}

\section{Study 1: Replicating Lab Experiments of PGG}

\subsection{Methodology}

To test whether LLMs can replicate lab experiments, we pick three lab experiments with human subjects. We create simulations for them, run it, and compare results to human behavior. These three studies focus on three different effects: (1) priming via game, (2) transparency of contributions, and (3) variation in endowment. They found the following:

\subsubsection{Experiment \#1} \label{oneofone} The first experiment studies the effect of positive and negative priming on contributions in the PGG. We obtain data on human behavior from a 2014 lab experiment in \textit{Judgement and Decision Making} \cite{Eriksson_Strimling_2014} with 100 human participants. Participants were explained the rules of the game, but told one of two different names for the game. \color{black} In the positive condition, people were told the game was called the "Teamwork Game." In the negative case, people were told the game was called the "Taxation Game." The remaining conditions were the same. In the positive priming condition ("Teamwork"), people on average contributed 60\% of their endowment. In the negative priming condition ("Taxation"), people on average contributed only 40\% of their endowment. \textbf{Thus, positive priming increases prosocial behavior.}


We test whether this effect applies to LLM-agents by simulating ten one-shot PGGs with four players, two under each priming condition and all four with a \$20 endowment, and comparing the average contribution amount of players from either group. Specifically, we prime players by including sentences containing the appropriate name in their private biographies (for example, "Alice is playing a game called the "Teamwork Game."). In the simulation, 1.6 times the amount of the public pool is split evenly amongst the players as their payoff - players are made aware of this to inform the way in which they act, but the simulation is of a one-shot PGG.

\subsubsection{Experiment \#2} The second experiment studies the effect of transparency of contributions in the PGG. We obtain data on human behavior from a 2017 review paper of 71 human lab studies published in \textit{Experimental Economics} \cite{transparencytwo}. \color{black} Specifically, the difference in contribution is presented for people who played the PGG in two conditions: one where people knew the contributions made by others and another where they did not. When there was transparency of contributions, people contributed 6\% more on average than when there was not. \textbf{Thus, transparency of contributions increases prosocial behavior.}



We test whether this effect applies to LLM-agents by simulating ten four-player one-shot PGGs, five with transparency of contributions and five without, and comparing the average contribution amounts in either condition. Specifically, we realize transparency by removing the moderation room - contributions are made in the "public" game room where all other agents can overhear other's contributions. In the simulation, 1.6 times the amount of the public pool is split evenly amongst the players as their payoff - players are made aware of this to inform the way in which they act, but the simulation is of a one-shot PGG.

\subsubsection{Experiment \#3} The third experiment studies the effect of unequal endowments on contributions in the PGG.  We obtain data on human behavior from a 2016 lab experiment published in \textit{Economics Letters} \cite{HARGREAVESHEAP20164}. \color{black}The experiment included four separate PGG games with different endowment conditions: (1) where all players had \$20, (2) where all players had \$50, (3) where all players had \$80, and (4) where one player had \$20, one player had \$50, and one player had \$80. The results of the experiment find that people endowed with \$20 and \$50 contributed roughly the same in both games were all players had the same endowment and games where endowments were varied, but that people endowed with \$80 contributed much more in games were all players had the same endowment than games in which endowments were varied.
\textbf{Thus, varying endowments does not seem to have an effect on prosocial behavior in "poor" and "medium" wealth individuals, but reduces prosocial behavior in "rich" individuals.}

We test whether this effect applies to LLM-agents by simulating twenty-five three-player one-shot PGGs, fifteen where players have equal endowments (five for \$20, \$50, and \$80 respectively) and ten where players have varied endowments (one player with \$20, one with \$50, and one with \$80), then comparing the average contribution of players with each endowment amount under the equal and varied endowment conditions. Specifically, we realize the variation in endowments simply by changing the players public biographies, where their endowments are set. In the simulation, 1.6 times the amount of the public pool is split evenly amongst the players as their payoff - players are made aware of this to inform the way in which they act, but the simulation is of a one-shot PGG.

\subsection{LLM-Agent Simulation Setup for the Public Goods Game}

\subsubsection{Multi-Agent LLM Architecture Implementation} 


When implementing the PGG in the multi-agent framework, we initialize it as follows. The moderator agent's name is "Moderator," and its public biography is simple: "This agent plays the role of the moderator". The moderator agent has no private biography. The moderator has an initial plan to start in the moderation room and wait for all players to make their contribution, and has instructions to then move to the game room and announce payoffs once all agents have made their contribution.

The player agents are given arbitrary alphabetical names (Alice, Bob, Casey, and David) and each have public biographies with a statement that the agent is playing a PGG-like game and contains their initial endowment, which all other agents can see. The player agents' private biography is blank by default, but is used to realize the priming condition of experiments  - for example, the name priming is realized by adding to the agents private biography, "You are playing a game called the \textit{Taxation Game}." \color{black} 

The architecture used does not have an explicit turn-taking process. \color{black} To implement the mechanics of turn-taking, player agents' have an initial plan to wait for the player before them to return from the moderation room. Player agents are ordered alphabetically by being told the player before them (in the case of Bob, Casey, and David) or by being told they are first (in the case of Alice). They also have instructions to move to the moderation room and make their contribution (with the payoff specified here for player agents to take into account) once the player before them has returned, and to not engage with any other player agents' in the game room until the Moderator comes to the room and payoffs are announced. In each \textit{agent loop}, the agent thus waits for the player before them to come back from the contribution room before going themselves, tells the moderator their contribution (which the moderator records to their memory), then returns to the main room.\color{black}

\subsubsection{Data Collection} As the simulation progresses, all of the thoughts and actions of each player agent (and the moderator agent) are displayed in separate output logs, each representing one of the agents involved in the simulation. Each output log contains a plethora of information, ranging from the events each agents observed, their reactions to them, and the creation of new plans, but we are only interested in the contributions they make.

 To collect data on the contribution amount of each agent, we used a script on each player agent's output logs. The script used string matching and regular expressions to look for when each agent was in the moderation room and the dollar amount said by the agent. This data was then output into a table. A human verified that the data was accurately transcribed by reading the sections identified by the script and making corrections as needed.

\subsection{Results}


\subsubsection{Does priming LLM agents via game name replicate the effect of priming humans via game name?}

\begin{figure}[h]
    \centering
    \includegraphics[width=0.45\textwidth]{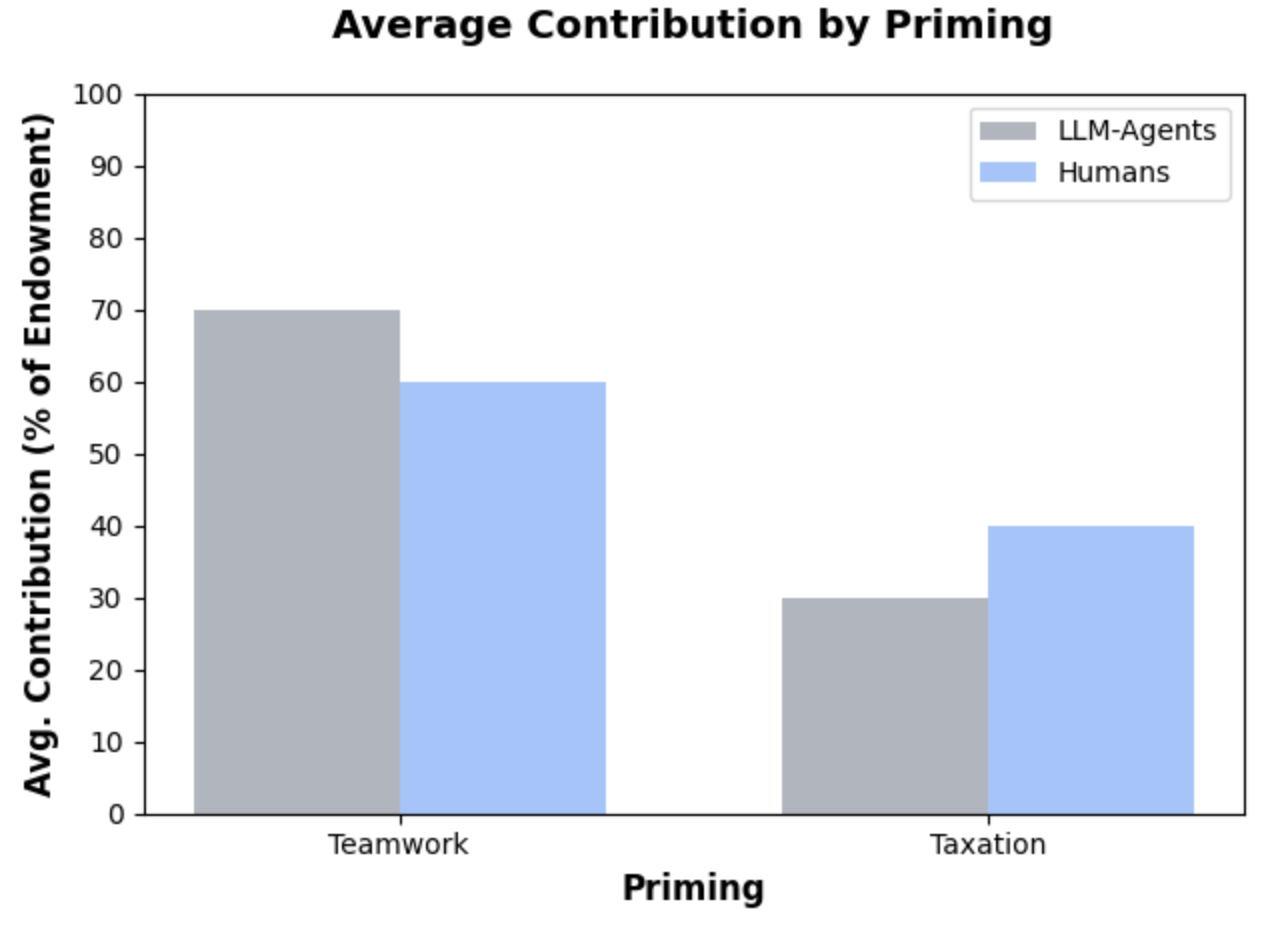}
    \caption{Average contributions for "Teamwork" and "Taxation" priming conditions in simulations with LLM-agents and experiments with human subjects. Average contributions are above 60\% in the "Teamwork" priming condition for both groups. Average contributions are below 40\% in the "Taxation" priming condition for both groups.}
    \Description{A bar chart titled "Average Contribution by Priming" compares the average contributions in two priming conditions: "Teamwork" and "Taxation." The y-axis represents average contributions as a percentage of endowment, ranging from 0 to 100\%. The x-axis lists the two priming conditions. Each condition has two bars: a gray bar representing LLM-agents and a blue bar representing human participants. In the "Teamwork" condition, both groups contribute above 60\%, with LLM-agents contributing slightly more than humans. In the "Taxation" condition, contributions are below 40\% for both groups, with humans contributing slightly more than LLM-agents.}
    \label{1Agraph}
\end{figure}


A t-test shows that the difference in LLM contributions between the two groups is statistically significant at the p < 0.01 level  
(\(t = 7.92, p < 0.00000001 \)). We use one-sample one-tail t-tests to compare the LLM-simulation sample data with the human average contribution, finding that under the "Teamwork" priming condition, the LLM contributions are greater than the human average of 60\% (\(t = 3.59, p = 0.001 \)) at the p < 0.01 level. Under the "Taxation" priming condition, LLM contributions are less than the human average of 40\% (\(t = -2.33, p = 0.016 \)) at the p < 0.05 level. \textbf{From these results, we conclude that priming LLM agents by presenting the PGG under different names replicates the direction of the effect of doing the same on humans, but may overestimate the magnitude.}\color{black} 

\subsubsection{Does introducing transparency in contribution between LLM agents replicate the effect of introducing transparency in contribution between humans?}

\begin{figure}[h]
    \centering
    \includegraphics[width=0.45\textwidth]{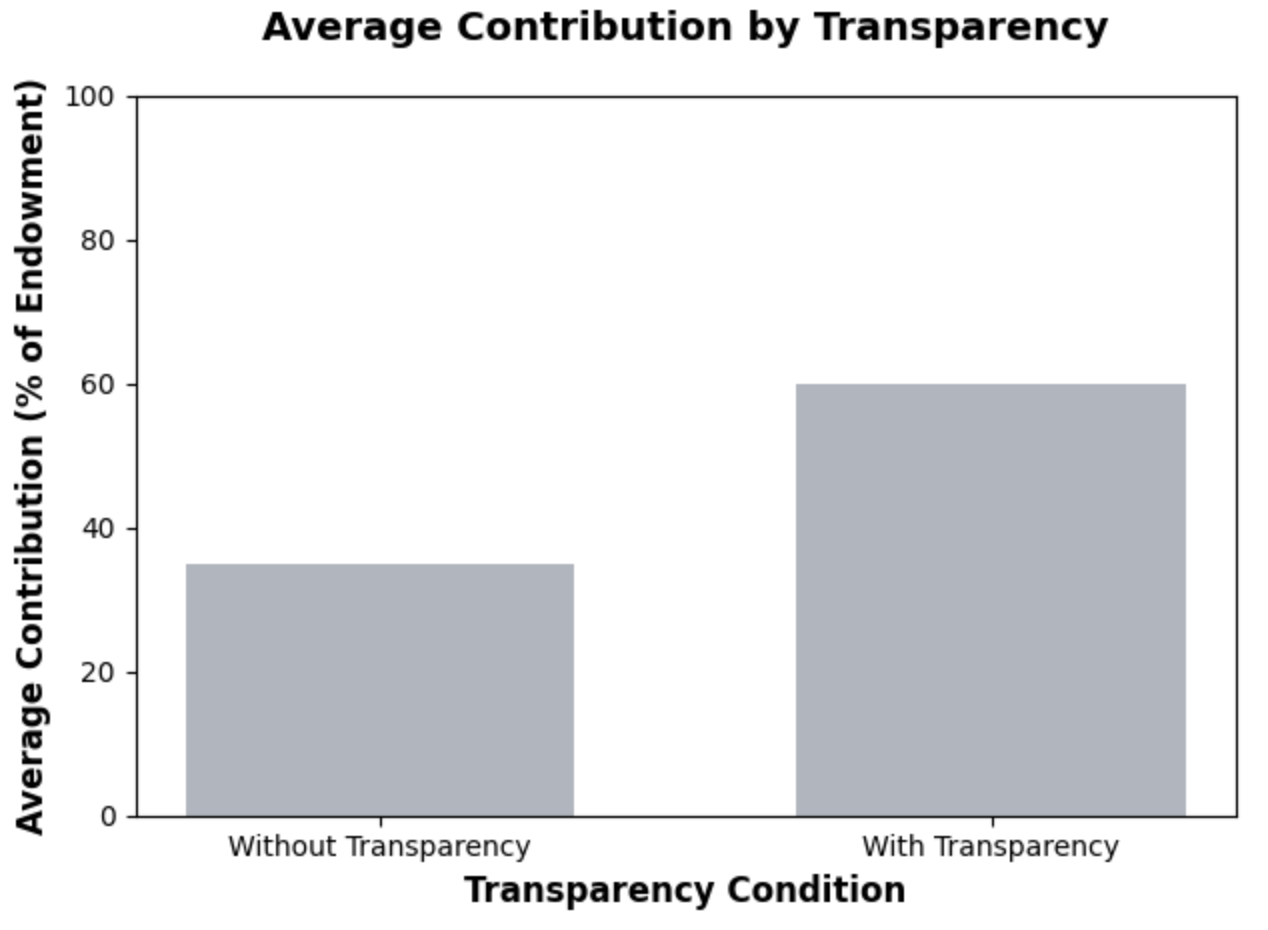}
    \caption{Average contributions with and without transparency of contributions for simulations. Experiments with humans show that average contributions in PGGs with transparency of contributions is 6\% higher than in PGGs without without transparency. In simulations with LLM-agents, average contributions in PGGs with transparency of contributions were 25\% higher than in PGGs without transparency. So, the direction of the difference is accurately captured.}
    \Description{A bar chart titled "Average Contribution by Transparency" compares the average contributions in two transparency conditions: "Without Transparency" and "With Transparency." The y-axis represents average contributions as a percentage of endowment, ranging from 0 to 100\%. The x-axis lists the two transparency conditions. The bar for "Without Transparency" is lower (35\%), indicating a lower contribution percentage, while the bar for "With Transparency" is noticeably higher (60\%). This suggests that contributions increase when transparency of contributions is present.}
    \label{1Bgraph}
\end{figure}

In experiments with human subjects, participants contributed on average 6\% more of their endowment under transparency conditions as compared to no transparency. \cite{transparencytwo}. Across five simulations of each condition (ten simulations total) on average, LLM agents playing the PGG with transparency contributed approximately 60\% of their \$20 endowment, whereas LLM agents playing the PGG without transparency contributed only approximately 35\% of their endowment. These results are in Figure \ref{1Bgraph}.

A one-tailed t-test shows that the mean contribution of LLMs under the transparency condition is significantly greater than the mean contribution of LLMs under the non-transparent condition (\(t = 2.23, p = 0.016 \)) at the p < 0.05 level. \color{black} \textbf{From these results, we conclude that introducing transparency of contributions in PGGs with LLM agents replicates the direction of the effect of doing the same on humans.}


\subsubsection{Does varying the endowments of LLM agents replicate the effect of varying endowments of humans?}

\begin{figure*}[t]  
    \centering
    \includegraphics[width=\textwidth]{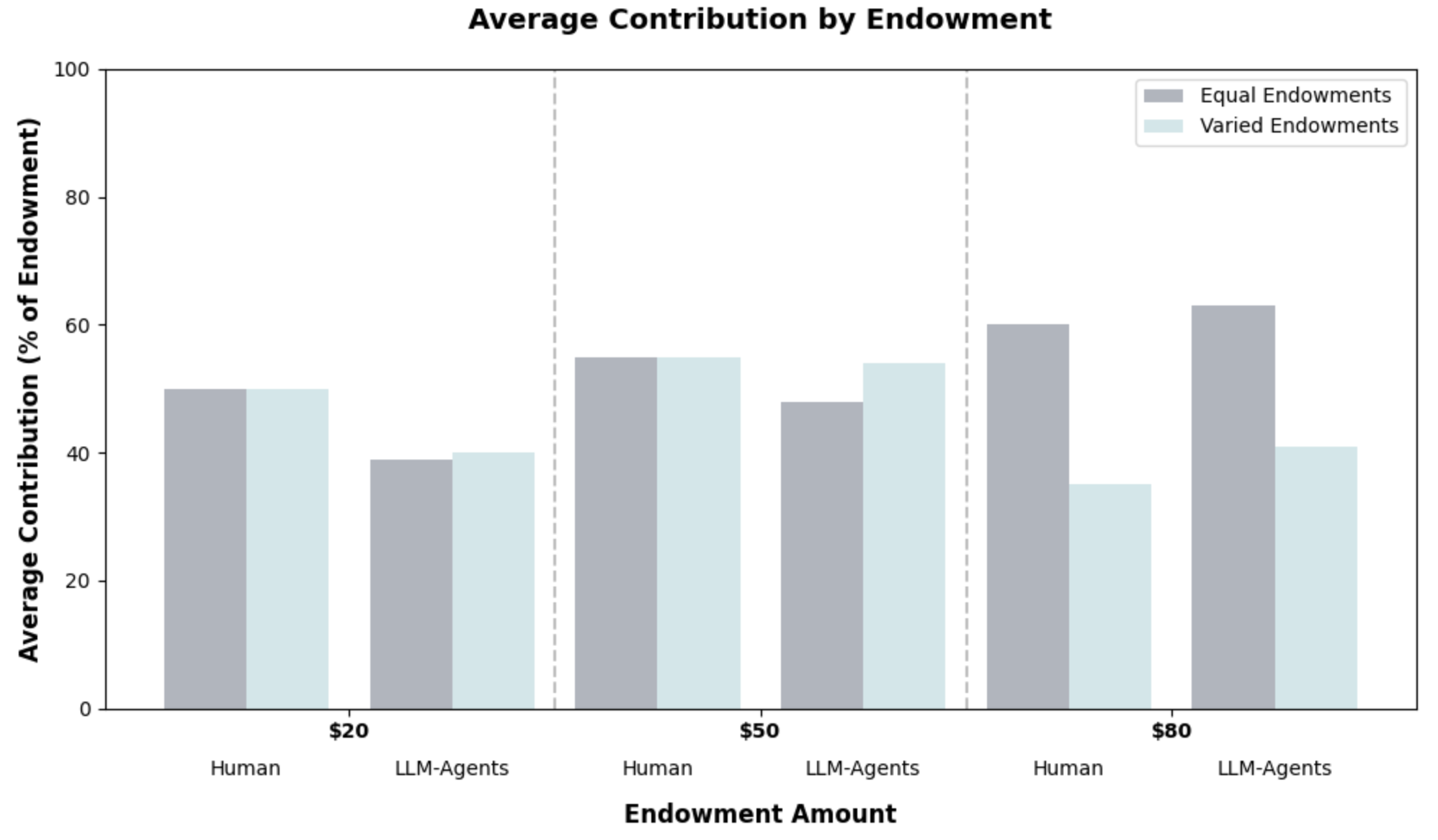}  
    \caption{Average contributions for equal and varied endowment conditions in simulations with LLM-agents and experiments with human subjects. Average contributions are roughly the same for humans and LLM-agents in the equal and varied endowment conditions with \$20 and \$50 endowments. Average contributions are lower in the varied condition than in the fixed condition with \$80 endowments for both humans and LLM-agents.}
    \Description{A bar chart titled "Average Contribution by Endowment" compares contributions as a percentage of endowment across different endowment amounts (\$20, \$50, and \$80) for humans and LLM-agents. The x-axis represents the endowment amounts and groups participants into "Human" and "LLM-Agents" under each endowment level. The y-axis represents the average contribution as a percentage of endowment, ranging from 0 to 100\%. Each group has two bars: a dark gray bar for "Equal Endowments" and a light blue bar for "Varied Endowments." Contributions generally increase with higher endowments, with variations between humans and LLM-agents. The impact of equal vs. varied endowments is also visible across conditions.}
    \label{1Cgraph}
\end{figure*}

In experiments with human subjects, it was found "poor" and "medium" wealth individuals contributed roughly the same amount in equal and varied endowment conditions, whereas "rich" individuals contributed more in the equal endowment condition than in the varied endowment condition \cite{HARGREAVESHEAP20164}. Across five simulations each of \$20-\$20-\$20, \$50-\$50-\$50, and \$80-\$80-\$80 equal endowment conditions, LLM agents on average contributed approximately 39\%, 48\%, and 63\% of their endowments, respectively. In five simulations of a \$20-\$50-\$80 varied endowment condition, on average the LLM agent given \$20 contributed approximately 35\%; the LLM agent given \$50 contributed approximately 42\%; and the LLM agent given \$80 contribute approximately 44\%. Thus, the contribution for agents endowed with \$20 was roughly the same between the equal and varied endowment conditions, whereas agents endowed with \$50 and \$80 both contributed less in the varied condition.These results are summarized in Table \ref{1Cgraph}.


T-tests shows that there is no statistically significant difference at the p < 0.1 between the contributions of LLM agents endowed with \$20 in the equal and varied endowment condition (\(t = -9.42, p = 0.678 \)), nor for the \$50 (\(t = -0.56, p = 0.58 \)) endowment - both results mirroring those of human subjects. However, a t-test shows that the difference between the contributions of LLM agents endowed with \$80 is statistically significant at the p < 0.1 level (\(t = -1.94, p = 0.065 \)). A one-tailed t-test specifically shows that the contribution of LLM-agents endowed with \$80 is significantly greater in the equal condition than in the varied condition (\(t = 2.05, p = 0.026 \)) at the p < 0.05 level. \color{black} \textbf{From these results, we conclude that varying endowments in the PGG with LLM agents replicates the effect of doing the same on humans.}



\section{Study 2: Testing Multi-Agent Systems Ability to Transfer Effects from non-PGG Lab Experiments to Simulations of the PGG}

To test whether LLMs can actually simulate human-like behavior and are not just parroting previously published lab experiments, we design two further experiments. These experiments take effects from lab experiments not containing the key word "public goods game." We then measure if LLMs are able to produce the expected outcomes in simulations of the PGG. We run two experiments; the first applies positive and negative priming methods used in other cooperation games, to assess if LLMs can generalize these effects to the PGG. The second tests priming effects from a one-shot PGG lab experiment, but over multiple rounds, incorporating the observed changes in priming effects over time from competition games. This approach confirms that the results in Study 1 were not simply due to LLMs parroting the results from a single lab experiment - it demonstrates that LLMs can integrate information from multiple various sources to simulate human-like behavior. 



\color{purple}





\color{black}

\subsection{Methodology}

\subsubsection{Experiment \#1} The first experiment uses a priming methodology used in a 2013 lab experiment of a modified solidarity game with 75 participants published in \textit{Plos One} \cite{moralsmatter}. The solidarity game is another cooperative economics game, but the priming methodology used \color{black} has not been tested on the PGG in prior lab experiments to our knowledge. Positive priming has shown to induce generosity in the PGG and other cooperative games, so we test whether LLMs are able to replicate the effects of a positive priming methodology used in a non-PGG lab experiment in simulations of the PGG. The positive priming involves showing participants three sentences alluding to "unity" ('\textit{we are family},' '\textit{mine is also yours},' and '\textit{caring for each other}'), while the negative priming involves showing participants three sentences alluding to "proportionality" ('\textit{how are you useful for me},' '\textit{I want to profit},' and '\textit{making a deal}'). The results of the lab experiment find that people in the positive priming condition displayed more generosity than people in the negative priming condition in the solidarity game. Thus, the positive priming increased prosocial behavior, meaning that it should increase contributions in the PGG as well.


We test whether this effect applies to LLM agents by simulating ten one-shot PGGs with four players, two under each priming condition, and all four with a \$20 endowment, and comparing the average contribution amount of players from either group. Specifically, we prime players by adding the three sentences to the player's private biography as something they have read (for example: "Alice read the following sentences prior to playing the game: '\textit{we are family},' '\textit{mine is also yours},' and '\textit{caring for each other}.'"). In the simulation, 1.6 times the amount of the public pool is split evenly amongst the players as their payoff - players are made aware of this to inform the way in which they act, but the simulation is of a one-shot PGGs.

\subsubsection{Experiment \# 2} The second experiment measures the effect of positive and negative priming over time (multiple rounds). Prior lab experiments of the one-shot PGG have shown that positive priming (by presenting the game as the "Teamwork Game") increases contributions while negative priming (by presenting the game as the "Taxation Game") decreases contributions \cite{Eriksson_Strimling_2014}\color{black}. Meanwhile, prior lab experiments of multi-round Bertrand competition games show that the effect of primings fades over time \cite{JIMENEZJIMENEZ201594}\color{black}. By combining these results, we expect that positive/negative priming should respectively increase/decrease the average contribution in the first round of the PPG, but should trend closer to half over multiple rounds. 


\color{black}



We test whether this effect applies to LLM-agents by simulating ten five-round PGGs with four players, two subject to a positive priming (the "Teamwork" priming discussed in Study 1, Experiment \#1) and two subject to a negative priming (the "Taxation" priming discussed in Study 1, Experiment \#1). We compare the average contribution in the first round for either priming condition with the average contribution in the last round for either priming condition. We also compare the initial contributions between priming conditions, checking that there was an initial disparity that ultimately begins to fade. In this experiment, payoffs are computed the same as in previous experiments (namely, 1.6 times the amount of the public pool split evenly amongst the players). Since there are multiple rounds, the payoff and amount not contributed are summed to determine each players endowment for the next round.

\subsection{Results}


\subsubsection{Does priming LLM agents with a methodology used in non-PGG cooperation game lab experiments have the expected result in LLM simulations of the PGG?}

\begin{figure}[h]
    \centering
    \includegraphics[width=0.45\textwidth]{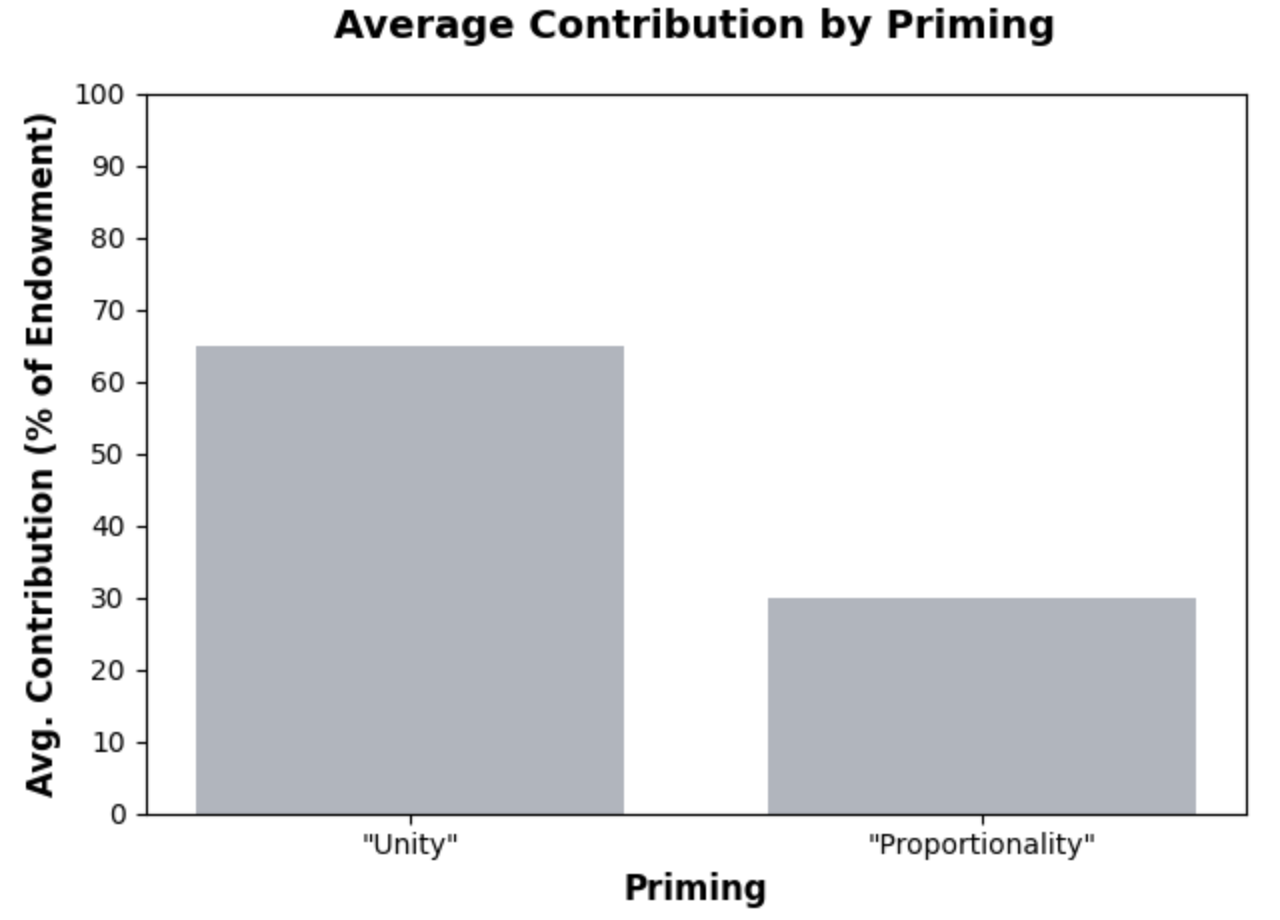}
    \caption{Average contributions for either "unity" or "proportionality" priming conditions. Average contributions under the "unity" condition are significantly higher than that under the "proportionality" condition.}
    \Description{A bar chart titled "Average Contribution by Priming" compares average contributions under two priming conditions: "Unity" and "Proportionality." The y-axis represents average contributions as a percentage of endowment, ranging from 0 to 100\%. The x-axis lists the two priming conditions. The bar for "Unity" is significantly higher, showing contributions above 60\%, while the bar for "Proportionality" is much lower, below 40\%. This indicates that priming with a "Unity" condition leads to higher contributions compared to the "Proportionality" condition.}
    \label{2Agraph}
\end{figure}

In experiments of other cooperation games, priming participants via sentences alluding to "unity" resulted in participants being more generous than participants primed via sentences alluding to "proportionality" \cite{moralsmatter}. In ten simulations, LLM agents primed with the sentences alluding to "unity" (Alice and Bob) contributed on average approximately 65\% of their initial endowment. In comparison, agents primed with the sentences alluding to "proportionality" (Casey and David) contributed on average only approximately 30\% of their endowment to the public pool. The results are summarized in Figure \ref{2Agraph}.

A t-test shows that the difference in contributions between the two groups is statistically significant at the p < 0.01 level (\(t = 8.76, p < 1\text{e}{-9} \)). \textbf{From these results, we conclude that the effects of priming on humans is transferred from papers not involving the PGG to simulations of the PGG with LLM-agents.}


\subsubsection{Does the effect of priming over time, observed to fade in non-PGG competition game lab experiments, hold in LLM simulations of the PGG?}
\begin{figure}[h]
    \centering
    \includegraphics[width=0.45\textwidth]{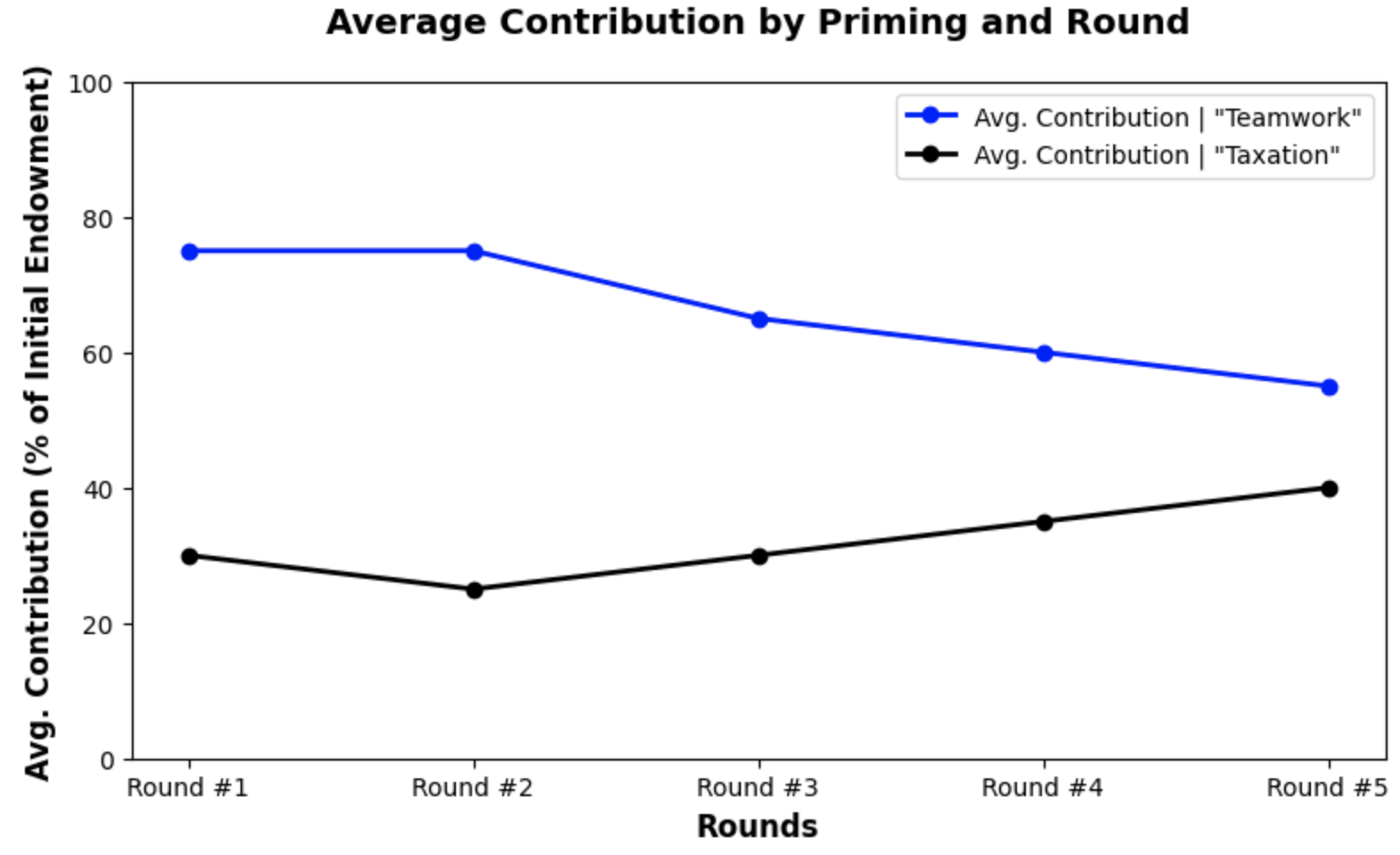}
    \caption{Average contributions in the all five rounds rounds for either "Teamwork" and "Taxation" priming conditions. Average contributions in the fifth round are closer to 50\% of the initial endowment, the average amount contributed without any priming \cite{labexperiments}. Hence, the effects of priming appears to fade over time in simulations with LLM-agents.}
    \Description{A line chart titled "Average Contribution by Priming and Round" shows how average contributions change over five rounds under two priming conditions: "Teamwork" and "Taxation." The y-axis represents average contributions as a percentage of the initial endowment, ranging from 0 to 100\%. The x-axis represents the round numbers (Round #1 to Round #5). The blue line represents contributions under the "Teamwork" priming condition, which starts high and gradually decreases over time. The black line represents contributions under the "Taxation" priming condition, which starts lower but increases slightly across rounds. By Round #5, contributions in both conditions converge closer to 50\%, suggesting that the effects of priming diminish over time in simulations with LLM-agents.}
    \label{2Bgraph}
\end{figure}

\bigskip
In experiments of competition games with human subjects, the effect of priming has been found to fade over time \cite{JIMENEZJIMENEZ201594}. In ten simulations, LLM agents presented a five-round PGG as the "Teamwork Game" (Alice and Bob) on average contributed 75\% of their initial endowment in the first round, compared to 55\% of their initial endowment in the fifth round. In comparison, agents presented the PGG as the "Taxation Game" (Casey and David) contributed 30\% of their initial endowment in the first round, compared to 45\% of their initial endowment in the fifth round. Hence, for both priming conditions, the average contribution was closer to 50\% of the initial endowment in the fifth round than in the first, meaning the effect of the priming was less pronounced in the fifth round than the first. The results are summarized in Figure \ref{2Bgraph}.

T-tests show that the difference in contribution amount between the rounds is statistically significant for the "Teamwork" priming condition at the p < 0.01 level (\(t = 4.32, p = 0.002 \)) and for the "Taxation" priming condition at the p < 0.05 level (\(t = -2.30, p = 0.047 \)). A t-test also shows that there is a statistically significant difference in the first round contributions of players in either priming condition (\(t = 9.64, p < 0.0000000001 \)). \textbf{From these results, we observe that the effect of priming LLM agents fades over time, concluding that LLMs are not just parroting past literature, but are able to transfer the multi-round effect observed in non-PGG lab experiments to simulations of the PGG.} The "Taxation" priming condition specifically suggests that the difference between first and fifth round contributions is not because contributions in PGGs with human subjects decrease over time \cite{labexperiments}, but rather that it is likely the consequence of a fading priming effect.


\section{Study 3: Simulating "In-the-Wild" Scenarios}



This study aims to test whether LLM-agents can demonstrate realistic prosocial-related behavior in "in-the-wild" real-world situations. Real-world situations require a greater level of complexity in setup than lab experiments, as individuals do not have as explicit of instructions on when and how to act or speak. Furthermore, there is a lack of specific papers for LLMs to draw from to simulate human behavior in such open-ended settings. We test two specific real-world situations: (1) a classroom setting and (2) a grocery store parking lot setting. In a classroom, we have anecdotal evidence from university faculty that changing the harshness of late policies or adding perturbations to student lives have an effect on student behaviors. Specifically, we test whether the emergent behavior of cheating, a "bad" form of cooperation, occurs in simulation. In leaving a grocery store parking lot, shoppers may be affected by other factors in their lives that change how they act \cite{scientificamerican2020shoppingcarts}. Specifically, we test whether the emergent behavior of returning their shopping cart, a form of prosocial behavior, occurs in simulation. We test if we can see these emergent behaviors in a basic simulation setup, finding that in both scenarios our basic setup is insufficient to observe the desired behaviors. Hence, we add a mechanism in each to increase the complexity of the setup that guides LLM simulations to produce results closer to expected human behavior.






\subsection{Classroom Setting}

\subsubsection{Setup}

\paragraph{Classroom Simulation Setup}

Towards simulating the behaviors in a classroom setting, we first try basic simulation setup with only one location, but find that expected student behaviors (specifically, cheating) are not present. Hence, we add a mechanism to the simulation setup - locations for private communication - to enable students to communicate with one another without the teacher overhearing and discuss the things that real students may discuss in private.

To simulate the classroom setting, we identify two main necessary components for the basic setup: (1) people, (2) a location, and (3) homework assignment and submission instructions. The simulation involves two types of people: one professor and multiple students (specifically, 3). These are all represented by separate LLM-agents, all located in the same location. Students are given various personality traits (over-achiever, procrastinator, or values work-life-balance) and can be affected by two perturbations (having a midterm or a particularly challenging assignment). The professor is instructed to announce the late-policy and make assignment announcements at a regular interval. Students are instructed to listen to assignment announcements at the interval at which the professor announces them, and to "work" on the assignments in the meantime. Students are able to converse with the professor and other students, although only by addressing all at the same time.

For the more complicated setup, we introduce using locations as a mechanism for private communication. The simulation requires three locations: (1) a classroom, where assignments are announced, (2) an office, where the professor goes while students "work", and (3) a work room, where students "work." All students know these locations exist; the students can move between all of them, while the professor only moves between the classroom and office. The professor is instructed to announce the late-policy and make assignment announcements at a regular interval in the classroom, and to remain in the office otherwise. Students are instructed to listen to assignment announcements in the classroom at the interval at which the professor announces them, and then to be in the work room otherwise, with the option to move to the office to talk to or ask the professor questions.


\paragraph{Multi-Agent LLM Architecture Implementation}

We implement the simulation by adapting the same architecture as in the previous studies. When implementing the classroom setting in the multi-agent framework, we initialize it accordingly. In both setups, the professor agent's name is "Professor." The professor agent's public biography contains their late policy for assignments, as well as a general sentence indicating that they are the instructor for the course. The professor agent has no private biography. In the basic setup, the professor has an initial plan to announce their late policy and the first assignment's due date. The professor agent has instructions to answer student questions and to announce additional assignments at regular intervals (five total). The student agents are given arbitrary alphabetical names (Alice, Bob, and Casey). The student agents' public biographies contain their personality types, whereas their private biographies contain the perturbations as needed. Student agents are given an initial plan to listen to the late policy and first assignment announcement. Student agents have instructions to: (1) "work" on assignments, (2) listen to the professor's announcements of new assignments, and (3) state how many days late they will need to submit each assignment.

After adding separate office, work, and class rooms, we modify the instructions to each agent to utilize the ability to move between them appropriately. The professor agent has an initial plan to start in the classroom and announce their late policy and the first assignment's due date, then to move to the office. The professor agent has instructions to (1) answer student questions in the office, (2) move to the classroom at regular intervals to announce additional assignments, and (3) never enter the work room. Student agents are given an initial plan to listen to the late policy and first assignment announcement in the classroom. Student agents have instructions to: (1) work on assignments in the work room, (2) return to the classroom at regular intervals to hear announcements for new assignments, (3) state how many days late they will submit each assignment, and (4) use the office to talk to or ask questions to the professor. 


\paragraph{Data Collection} As the simulation progresses, all of the thoughts and actions of each student agent (and the
professor agent) are displayed in separate output logs, each representing one of the agents involved in the simulation.
Each output log contains a plethora of information, ranging from the events each agents observed, their reactions to
them, and the creation of new plans, but we are only interested in tracking the specific behavior of cheating, which occurs by LLM-agents requesting to see and/or copy other's work. To collect data on the presence of this behaviors, we manually read through each student agent's output logs. A human manually recorded whether or not cheating was suggested and formatted this information into a table.

\subsubsection{Methodology}

We run simulations with three different late policies: (1) a lenient late policy (LLP) in which assignments turned in late are not given any penalty, (2) a medium late policy (MLP) in which assignments are docked 10\% for each late day, and (3) a harsh late policy (HLP) in which late assignments are not accepted. We also run simulations under three different perturbation conditions: (1) where students do not have any perturbations (P0), (2) where students have a midterm during the third assignment period (P1), and (3) where students have a midterm during the third assignment period and the second assignment is especially challenging (P2). We ran five simulations of each of the nine combinations of late policy and perturbation conditions (LLP-P0, LLP-P1, LLP-P2, MLP-P0, MLP-P1, MLP-P2, HLP-P0, HLP-P1, and HLP-P2) under two conditions: with or without locations. In each simulation, we recorded if any student proposed cheating at any point.

\subsubsection{Results}

\begin{figure*}[t]  
    \centering
    \includegraphics[width=\textwidth]{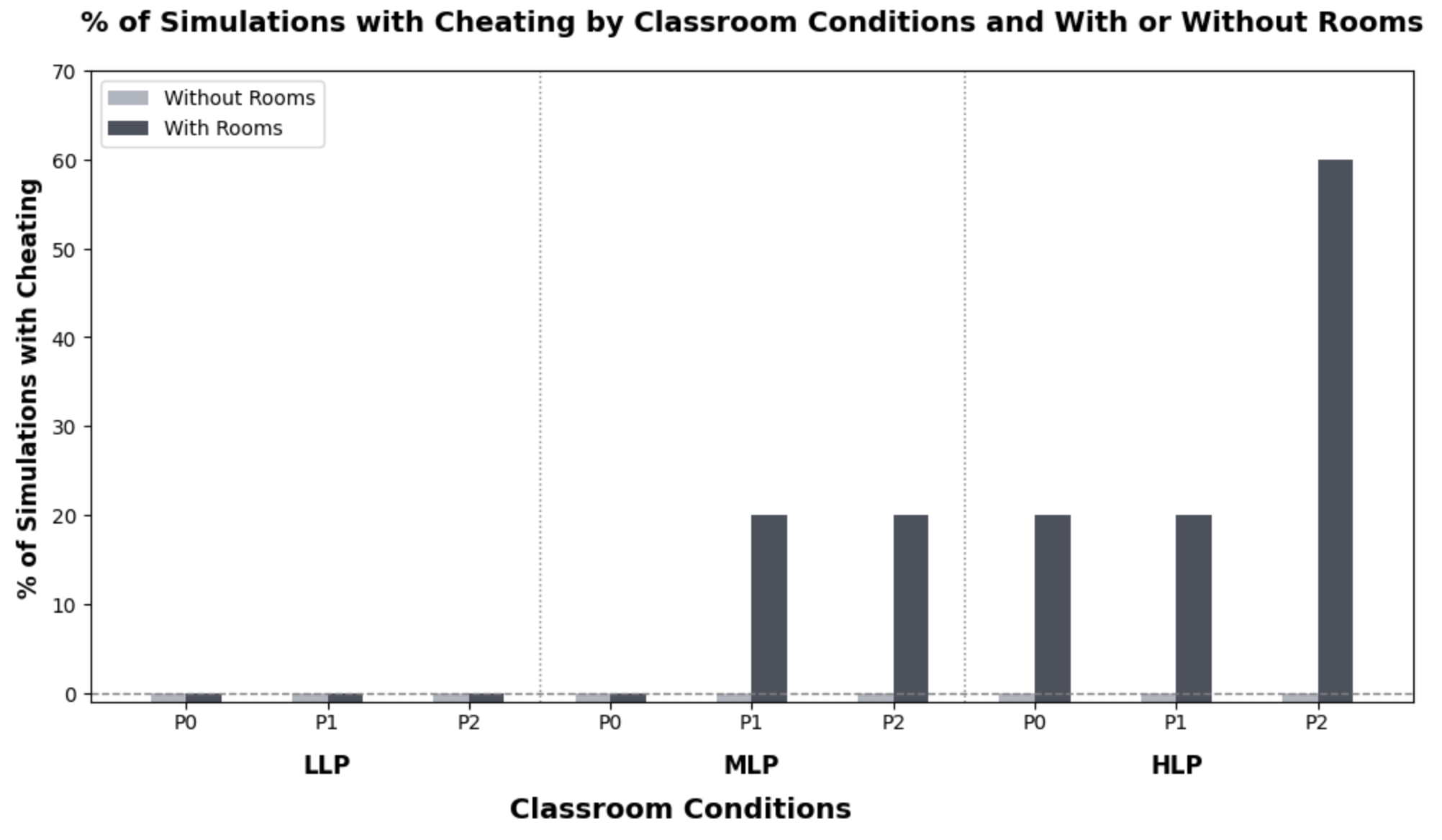}  
    \caption{Percentage of simulations in which cheating was observed under the 9 late policy and perturbation conditions, for simulations with and without multiple locations. Without multiple locations, cheating is never observed. With multiple locations, cheating first begins to be observed under the medium late policy under the one perturbation condition. The number of simulations in which cheating occurs is the same for the medium late policy in the two perturbation condition, as well as for the harsh late policy under the no perturbation and one perturbation conditions. The number of simulations in which cheating occurs is greatest with the harsh late policy under the two perturbation condition.}
    \Description{A bar chart titled "\% of Simulations with Cheating by Classroom Conditions and With or Without Rooms" illustrates the percentage of simulations in which cheating was observed under different classroom conditions. The y-axis represents the percentage of simulations with cheating, ranging from 0\% to 70\%. The x-axis represents three classroom conditions: LLP (Low Late Policy), MLP (Medium Late Policy), and HLP (Harsh Late Policy), each divided into three perturbation conditions (P0, P1, P2). Two sets of bars are shown for each condition: dark gray for "Without Rooms" and light gray for "With Rooms." Results indicate that cheating is not observed under LLP regardless of perturbation conditions. Under MLP, cheating emerges in P1 and P2, with slightly higher occurrences when multiple locations (rooms) are present. Under HLP, cheating is observed across all perturbation conditions and increases sharply in P2, particularly when multiple locations are involved, reaching the highest cheating percentage in the entire dataset.}
    \label{3Agraph}
\end{figure*}

In simulations with only one room, we found that students never suggested cheating, likely due to the professor always being in the same room as them. However, the introduction of locations changed this. In the lenient late policy condition, students still never suggested cheating, regardless of perturbations. In the medium late policy, students never suggested cheating in the no perturbation condition, but did so in one simulation in each of the one perturbation and two perturbation conditions. In the harsh late policy, students suggested cheating in all three perturbation conditions: students suggested cheating at an equal frequency in the no perturbation and one perturbation conditions, and the most in the two perturbation condition. The results are presented in Figure ~\ref{3Agraph}. \textbf{From these results, we conclude that locations can be used a mechanism for private communication to observe emergent behaviors such as cheating (a "bad" form of cooperation), which can only occur if students are able to converse without the professor overhearing.}


\subsection{Shopping-Cart Return Setting}

\subsubsection{Setup}

\paragraph{Shopping-Cart Return Simulation Setup}

There are three main necessary components for simulating the shopping-cart return setting: (1) people, (2) locations, and (3) instructions on what to act on. The simulation involves one person, a shopper, who is preparing to leave the parking lot and represented by a LLM-agent. The simulation requires two locations: (1) the area where the shopper has parked, and (2) the receptacle, where carts \textit{should} be returned. The shopper agent knows these locations exist, and can decide whether or not to move between them. The shopper agent can be affected by conditions that affect their ability to return their cart (being far from the receptacle or having a child with them) \cite{scientificamerican2020shoppingcarts}. The shopper is told to prepare to leave the parking lot and instructed that they cannot leave with their shopping cart - the shopper hence must decide \textit{something} to do with their shopping cart.

\paragraph{Multi-Agent LLM Architecture Implementation} We implement the simulation by adapting the same architecture as in the previous studies. When implementing the shopping-cart return setting in the multi-agent framework, we initialize it accordingly. The shopper agent is given the name "Shopper." The shopper agent has a blank public biography. The shopper agent's private biography states that they have a cart, and conditions that affect their ability to return their cart. The shopper agent has an initial plan to prepare to leave the parking lot, then instructions to act on the outcomes of their initial plan. We create two locations: (1) which represents the area where the shopper has parked and (2) the designated shopping-cart return receptacle. This way, we capture that LLM-agents must move in order to return their shopping cart.

However, we find that simply stating the condition affecting the shopper's ability to return their cart ("you are far from the receptacle," and "you have a child") is insufficient to have the desired effect of shoppers returning their carts less frequently; the prompt must allude towards what is at stake ("you are \textit{parked across the parking lot} from the receptacle," and "you have a \textit{five-month old} child"). In the condition where shoppers are far from the receptacle, adding that shoppers are "\textit{parked across the parking lot}" alludes to the fact that going to receptacle will take effort, whereas adding that the child is an infant alludes to the fact that the child cannot be left unattended. We present this as stake-prompting (SP), a mechanism to have prompts allude to what is at stake, using which LLM-agents can be guided towards demonstrating more human-like behavior in simulation.

\paragraph{Data Collection} As the simulation progresses, all of the thoughts and actions of the shopper agent are displayed in an output log. The output log contains a plethora of information, ranging from the events the agent observed (which is minimal in this simulation), their reactions to them (similarly minimal), and the creation of new plans. But we are only interested in recording whether or not the shopper agent returns their cart, based on whether the shopper agent moves to the receptacle location. A human manually recorded this information and formatted it into a table.

\subsubsection{Methodology} We run simulations with two different conditions that shoppers can be affected by: (1) being far from the receptacle (FFR) and (2) having a child (HAC). We run five simulations each of either condition under with and without stake-prompting. Given that the condition affecting the shopper should influence them to not return their shopping cart, we track whether or not the shopping cart is returned in each simulation and then compare the frequency of returns between simulations with and without stake-prompting.

\subsubsection{Results} 
\begin{figure}[h]
    \centering
    \includegraphics[width=0.45\textwidth]{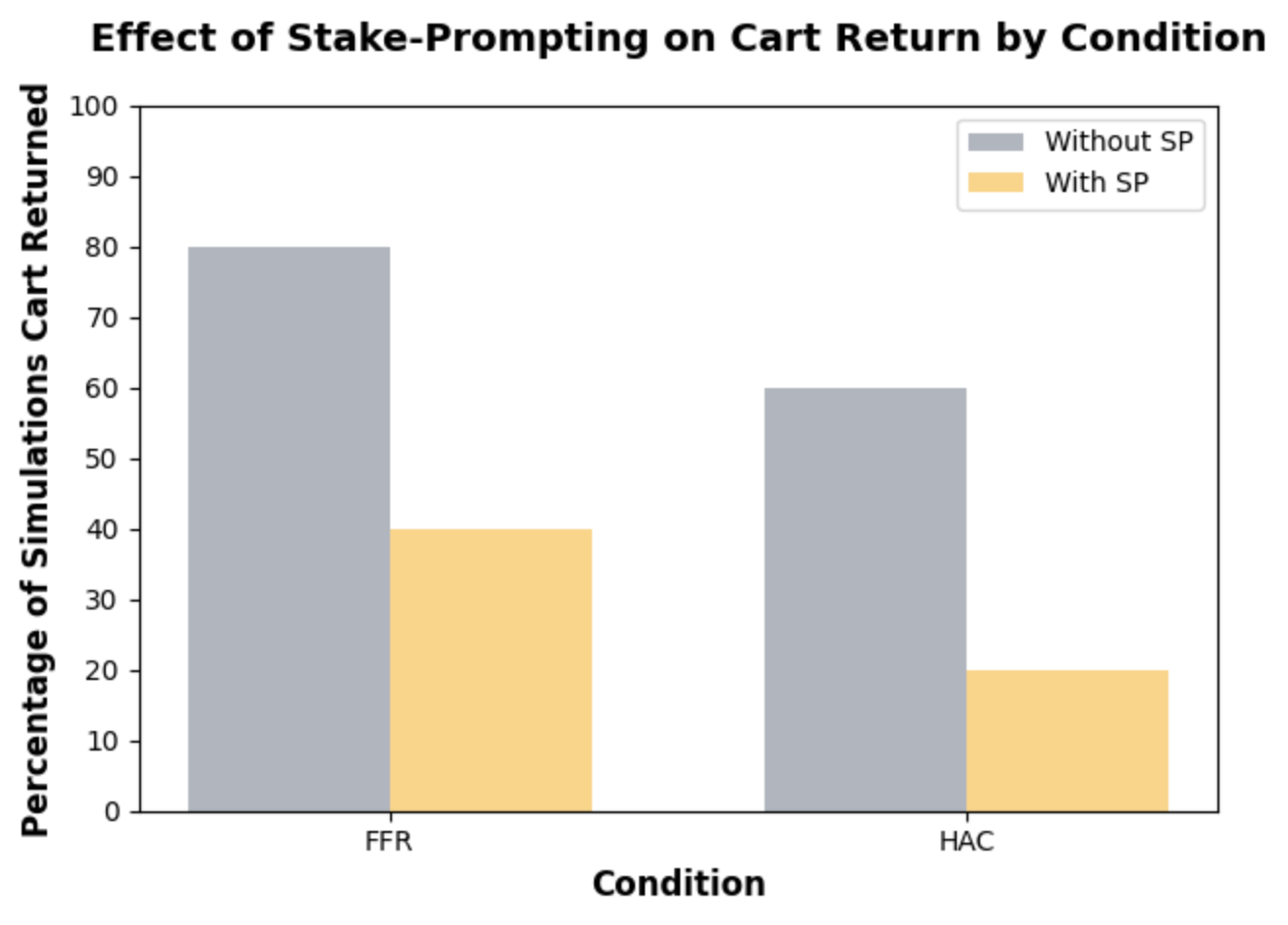}
    \caption{Percentage of simulations in which the shopper agent decided to their shopping cart for both external conditions tested and under either prompting condition: with or without stake-prompting. LLM-Agents returned their cart less frequently with stake-prompting than without.}
    \Description{ bar chart titled "Effect of Stake-Prompting on Cart Return by Condition" illustrates the percentage of simulations in which a shopper agent returned their shopping cart under two conditions: FPR and HAC. The y-axis represents the percentage of simulations where the cart was returned, ranging from 0\% to 100\%. The x-axis lists the two conditions. Each condition has two bars: a gray bar for the "Without SP" (stake-prompting) condition and a yellow bar for the "With SP" condition. Results show that cart return rates are higher when no stake-prompting is applied, with significantly lower return percentages when stake-prompting is present. This pattern is consistent across both conditions, suggesting that LLM-agents return their carts less frequently when stake-prompting is introduced.}
    \label{3Bgraph}
\end{figure}

In simulations without stake-prompting, we find that shopper agents still frequently return their carts despite the conditions affecting their ability to do so. However, alluding to what is at stake as a result of the condition  causes shopper agents to stop returning their carts as often. The results are summarized in Figure \ref{3Bgraph}. \textbf{From these results, we conclude that alluding to what is at stake via stake-prompting causes LLM-agents to demonstrate less prosocial behavior when affected by personal conditions, like humans.}

%% file: 5_discussion.tex
\section{Discussion}
\subsection{Generalizable Mechanisms to Enable Unbounded Behaviors in Simulation}

Simulations of real-world scenarios are particularly valuable when they produce realistic human behaviors that were not explicitly prompted. We refer to these as "unbounded behaviors" because they occur outside the expected framework of the simulation. Such behaviors are difficult to predict with traditional models, but multi-agent LLM systems have the unique potential to generate them. In Study 3, we observed unbounded behaviors such as cheating in a classroom simulation and shoppers neglecting to return carts in a parking lot.

However, in these scenarios, additional mechanisms were needed to enable the emergence of unbounded behaviors. For instance, we had to introduce private rooms to observe cheating in the classroom, and we had to add clear stakes (such as "you have an infant waiting in the car") to see agents failing to return shopping carts. 
If every simulation requires a custom mechanism to produce specific unbounded behaviors, the value of simulations diminishes —they would only reflect behaviors that researchers expect, thus undermining the purpose of simulation. However, if the mechanisms are generalizable, they can be widely implemented, allowing simulations to still produce meaningful unbounded behaviors. The question remains: were the mechanisms we added general or specific? With only two examples, it's hard to draw definitive conclusions, but there is reason to believe they could be broadly applicable.

In the classroom simulation, the introduction of private rooms was necessary to observe cheating. In hindsight, it seems obvious that students would only cheat if they had a private space to communicate without getting caught. Broadly speaking, communication is a fundamental mechanism in society that enables many unexpected outcomes. The invention of the telephone, the internet, and even home-delivered mail has had significant effects on the economy (increasing global trade and commerce), employment (enabling remote work), and culture (connecting distant people). Any simulation that tests cooperation will likely need diverse communication channels (private, public, synchronous, asynchronous) to allow for various forms of interaction and collaboration.

In the shopping cart simulation, adding stakes was essential. Studies show that people with a child waiting in the car are less likely to return their cart. However, we initially did not see this effect in the simulation. Upon reflection, the condition "having a child" was too vague. If the child was 12 years old, there was little pressure to skip returning the cart. But if the child was 5 months old, there was significant urgency. When we clarified the child's age in the simulation, the unbounded behavior emerged for parents with a 5-month-old more frequently than for parents with a child of unspecified age. This outcome makes sense, as it was a flaw in the initial setup to specify only "having a child" without including age. To simulate unbounded behaviors effectively, simulations need enough contextual details for agents to understand the stakes and reason for their actions.

In the future, more work is needed to discover a set of mechanisms that enable simulations to exhibit rich, bounded human behavior. Communication channels and detailed contextual stakes are likely two such mechanisms, but there are probably many more that need to be discovered.

\subsection{How can policymakers benefit from imperfect simulations with multi-agent systems?}
These experiments showed that multi-agent LLM simulations can replicate the direction of effects (though perhaps not the magnitude) and extend to situations not explicitly covered in the literature. They also have the capability to exhibit unbounded actions. These are all encouraging signs that these simulations could be a valuable tool for policymakers, but the reality is that they will never be perfect. The question is: can they still be useful despite their imperfections?

Even if simulations are not completely accurate, they can still serve as ideation tools to imagine different possible outcomes. Prospection — the ability to think and reason about the future—is a hallmark of human intelligence, yet it is something that many people struggle with. When thinking about the future, people often find it difficult to account for the complexity of multiple independent actors' decisions, fall prey to cognitive biases (such as optimism bias when they want a policy to succeed), and struggle to imagine novel situations for which they have little prior experience. Although future thinking is important, it is cognitively taxing, and any aids might be better than relying solely on our limited cognition. LLM systems, while imperfect, are relatively fast, low-cost, and can run multiple scenarios. In many cases, a reasonable simulation is better than none at all.

If simulations are accurate in predicting the direction of effects and relative effect sizes (even if not their exact magnitude), they can still be valuable for testing different policies to determine whether they will be effective (even if not precisely how effective). This can be especially useful because some policies are known to fade quickly over time, becoming ineffective unless frequently renewed, while others persist by establishing new norms—such as making organ donation the default option instead of requiring an opt-in. A simulation, particularly one that tracks effects over time, could reveal these policy limitations. Additionally, simulations can expose potential negative externalities, such as added bureaucracy or unintentionally excluding participants (for example, when new market regulations make it harder for newcomers to enter).

Simulations can also serve as a boundary object that helps individuals from diverse backgrounds integrate their knowledge. Many complex scenarios involve various areas of expertise and perspectives. In a city, a proposal like introducing a bike lane affects many departments —- transportation, waste management, emergency route planning, and legal services. Instead of each department prioritizing its own concerns, a simulation might help different groups see how their concerns intersect, fostering a more collaborative discussion. Thus, simulations could function as a social glue for large organizations.

While there is much work to be done with policymakers to determine the value of imperfect simulations, there are reasons to believe they could be useful for tackling a challenging and important problem that currently has few affordable and accessible feedback mechanisms.

\subsection{Future interfaces for simulations}
Although the simulations in this study were set up and run in a console without a specialized user interface, there are many ways people might want to interact with simulations in the future that would require rich user interfaces. After a simulation has been run, users may want to step through it to better understand the sequence of events that led to a particular outcome. They might also wish to compare two runs of the simulation with different outcomes to hypothesize which factors contributed to those differences. While these simulations were run in a batch mode (without user interaction), many users may prefer interactive simulations that allow them to interrupt the process, adjust parameters, or intervene to try to create or prevent certain outcomes. Additionally, users may want to fork a simulation at a particular stage to explore multiple parallel scenarios.

Since designing and validating a comprehensive simulation environment with all the necessary mechanisms is challenging, there could be a future where people use or modify existing environments rather than building one from scratch. For example, a large city simulation could be utilized by various organizations: city governments could use it to prepare for natural disasters, hospitals to allocate resources during medical emergencies, businesses to explore store locations, and civic planners to mitigate the effects of climate change. Such environments could be developed collectively and made available to the public, much like open-source code. However, there are also ethical concerns; simulations could be misused in harmful ways, and there may be reasons to restrict access to simulations of public places or sensitive areas.  As LLM simulations become easier to use, it is critical to build safegaurds to limit "bad actors" from using simulations to model nefarious situations like war or voter/consumer manipulation. Furthermore, overreliance on simulations is a major concern - AI simulations are much cheaper and easier to conduct than real design and policy work. However, when designing policies for people, it is still important to run experiments with real humans and use traditional methods to think through all possible consequences; simulations should only be used as an additional thinking tool, as opposed to being viewed as a standalone representation of ground truth. \color{black}

\section{Limitations and future work}
This paper represents an initial exploration of multi-agent LLM systems as a tool for simulating human behavior and has several limitations. While the findings demonstrate that the proposed approach is feasible, there remain many open questions and challenges that need to be addressed.

This paper focused on prosocial cooperative behavior and used the public goods game (PGG) as a vehicle for many of the preliminary studies (Studies 1 and 2). Although the PGG closely mimics real-world scenarios like taxation and charitable donations, it has limitations. First, the outcomes of the PGG depend on its multiplication factor for increasing the public pot. Typically, a factor of 1.6 is used, but if the multiplication factor is lower, the potential benefit is reduced, which could alter the game's dynamics. We lack human data for all possible values of the multiplication factor for comparison. Additionally, there are many variants of the PGG that were not tested: for example, punishment is an important mechanism that was not explored, though it has been shown to replicate in other games like the Ultimatum Game. There are countless mechanisms to replicate, and the more games we replicate, the more confidence we can have that multi-agent LLM systems are consistent with human behavior. At this point, there is a growing body of evidence from AI, HCI, economics, and psychology suggesting that LLM-agents can capture much of human behavior; what remains is to identify and refine the mechanisms that enable successful simulation.

Our second study, along with other studies replicating multiple unpublished psychology experiments ~\cite{hewitt2024predicting}, indicates that LLMs are not simply echoing the results of previous papers - there is potential for LLMs to simulate human behavior in novel scenarios. It is unclear how LLMs possess these abilities since they don't have inherent motivations like humans. Perhaps \color{black} they are using knowledge from past experiments and transferring principles learned from one context to new situations. An extreme hypothesis is that LLMs may not be relying on experimental knowledge at all but have genuinely understood aspects of human nature from their training data. However, the limits of these abilities are also uncertain. For any novel scenario in a simulation, we cannot be certain whether the LLM is capable of simulating the required behaviors. While previous work has shown that not all studies replicate, future research will need to test the boundaries of when LLMs are more or less likely to simulate human behavior accurately.

Our studies primarily tested single interventions, but real-world scenarios are more complex, often involving multiple policies that can interact in unpredictable ways. For example, adding transparency might increase donations in the PGG, while negative priming might decrease them. If both policies are present, the outcome could vary depending on the specifics of the situation, and there may not be a definitive ground truth to compare against. In the future, more psychological experiments may need to be conducted to inform simulation design.

Much of the experimental evidence from the social sciences is based on Western, Educated, Industrialized, Rich, and Democratic (WEIRD) populations. If LLMs do replicate human behavior, it is uncertain whether they can replicate all human behaviors, or only those represented in training data, such as experiments and online content. Further research is needed to understand to what extent LLMs can simulate behaviors of diverse groups, including children, marginalized communities, and others who are underrepresented in the training data. Past research has demonstrated that people from different types of communities act differently in the same strategic scenarios \cite{oosterbeek2004cultural, henrich2000does, kozitsina2020ethnicity} - it is important to ensure that simulations are representative of the population for which policymakers using them are designing for. \color{black} Additionally, people often behave differently toward their in-group versus an out-group, and it is unclear whether LLMs can capture these nuances.

The foundational LLMs used in these simulations continue to evolve. Simulations that replicate human behavior today may not do so in the future. As LLMs are trained on new datasets, their behavior could change for better or worse; high-quality human behavior data (e.g., from video sources) could improve the model, while low-quality or synthetic data could lead to degradation (e.g., mode collapse). Reinforcement Learning from Human Feedback (RLHF) can also cause certain LLMs, such as GPT-4, to display hyper-rational behaviors that exceed typical human responses, especially in emotionally charged situations where humans would behave more irrationally. Experimental results show that LLMs may react to induced emotions (e.g., happiness or anger) in disproportionately rational ways~\cite{mozikov2024good}. Future work may need to track how well simulation results replicate on different models over time.


Lastly, there is a gap between how role-playing tasks are simulated and actual cognitive processes. Naive priming in role-playing situations often leads to caricature-like behaviors instead of genuine cognitive modeling. This is because the underlying RLHF mechanism relies on human crowd workers providing feedback on what ~\emph{they think seems like} plausible behavior for a particular character in a particular situation. However, stereotypical perceptions of behavior can often not match real human behavior in moral dilemmas. For instance, many people may think a Buddhist monk would react to the trolley problem by avoiding action, when in reality surveys have shown the opposite. Moreover, the humans in the loop tend to belong to similar populations characterized as “poorly paid gig workers” ~\cite{casper2023open}, biasing the kinds of perceptions that are represented. Addressing these limitations may involve leveraging techniques like Bayesian induction and internal dialogue to foster more sophisticated representation of agent behaviors.

%% file: 6_conclusion.tex
\section{Conclusion}

This study demonstrates the potential of multi-agent LLM systems to simulate cooperative prosocial behavior, particularly in the context of public goods games. Our findings show that these systems can not only replicate the directional effects observed in human public goods game experiments but also apply priming effects from non-public goods game experiments and incorporate the fading of priming effects over time, as observed in multi-round games. The ability of LLMs to transfer effects across different contexts suggests a deeper understanding of human behavior rather than mere recapitulation of specific experimental results. Our research also highlights the need for additional mechanisms, such as private communication channels and stake-prompting, to accurately capture the nuances of human behavior in real-world scenarios.

While these results are promising, they also underscore the importance of careful implementation and interpretation when using LLM simulations to inform policy decisions. The observed exaggeration of effect magnitudes and the potential for bias inherent in LLM training data necessitate a cautious approach. Future research might focus on refining these simulation techniques, addressing limitations, and developing robust frameworks for validating LLM-based behavioral predictions against real-world outcomes. In sum, these studies suggest that LLM-based simulations, when used judiciously and with an understanding of their limitations, can become a valuable tool for policy designers, offering insights into potential human reactions to policies whose effects may be both difficult to anticipate and crucial to understand.